\documentclass[aps,prc,twocolumn,superscriptaddress,showpacs]{revtex4}
\usepackage{graphicx}
\usepackage{dcolumn}
\newcommand{\beq}{\begin{equation}} 
\newcommand{\eeq}{\end{equation}} 
\newcommand{\ds}{\displaystyle} 
\newcommand{\beqar}{\begin{eqnarray}} 
\newcommand{\eeqar}{\end{eqnarray}} 
\begin{document} 
 
\title{Violation of energy-per-hadron scaling in a resonance matter}

\author{L.~Bravina}
\altaffiliation[Also at ]{Institute for Nuclear Physics, Moscow 
State University, RU-119899 Moscow, Russia
\vspace*{1ex}}
\affiliation{Institut f\"ur Theoretische Physik, Universit\"at 
T\"ubingen, Auf der Morgenstelle 14, D-72076 T\"ubingen, Germany
\vspace*{1ex}}
\author{Amand~Faessler}
\affiliation{Institut f\"ur Theoretische Physik, Universit\"at 
T\"ubingen, Auf der Morgenstelle 14, D-72076 T\"ubingen, Germany
\vspace*{1ex}}
\author{C.~Fuchs}
\affiliation{Institut f\"ur Theoretische Physik, Universit\"at 
T\"ubingen, Auf der Morgenstelle 14, D-72076 T\"ubingen, Germany
\vspace*{1ex}}
\author{Zhong-Dao Lu}
\affiliation{China Institute of Atomic Energy,  
P. O. Box 275(18), Beijing 102413, China
\vspace*{1ex}} 
\affiliation{China Center of Advance Science and 
Technology (CCAST), Beijing 100080, China
\vspace*{1ex}} 
\author{E.E.~Zabrodin}
\altaffiliation[Also at ]{Institute for Nuclear Physics, Moscow 
State University, RU-119899 Moscow, Russia
\vspace*{1ex}}
\affiliation{Institut f\"ur Theoretische Physik, Universit\"at 
T\"ubingen, Auf der Morgenstelle 14, D-72076 T\"ubingen, Germany
\vspace*{1ex}}

\date{\today} 
 
\begin{abstract} 
Yields of hadrons, their average masses and energies per hadron
at the stage of chemical freeze-out in (ultra)relativistic heavy-ion
collisions are analyzed within the statistical model. The violation
of the scaling $\langle E \rangle / \langle N \rangle \cong 1$\,GeV 
observed in Au+Au collisions at 
$\sqrt{s} = 130${\it A} GeV is linked to the formation of 
resonance-rich matter with a considerable fraction of baryons and 
antibaryons. The rise of the energy-per-hadron ratio in 
baryon-dominated matter is discussed. A violation of the scaling 
condition is predicted for a very central zone of heavy-ion 
collisions at energies around 40{\it A} GeV.
\end{abstract}
 
\maketitle 

The properties of nuclear matter under extreme conditions have been
the subject of intensive experimental and theoretical studies during 
the last few decades. Up to now experiments with heavy-ion collisions 
remain the only means to explore the properties of hot and dense
nuclear matter in the laboratory. Two energy ranges connected with 
predicted phase transitions in the nuclear medium have been studied 
especially vigorously, - the intermediate energy range, where the 
nuclear liquid-gas phase transition might occur, and the range of 
relativistic and ultra-relativistic energies, where the phase 
transition to a deconfined quark-gluon plasma (QGP) and a restoration 
of chiral symmetry should take place (see \cite{QM99,QM01} and 
references therein). The search for the QGP formation is one of the 
top-priority goals of the heavy-ion collider program at Relativistic 
Heavy Ion Collider (RHIC) in Brookhaven, which is operating since 
1999, and at the forthcoming Large Hadron Collider (LHC) at CERN.

In order to reveal possible fingerprints of the QGP formation, various 
microscopic (transport, string, cascade) and macroscopic (thermal and 
hydrodynamic) models are widely used for the analysis of measured 
particle abundances and energy spectra. In macroscopic scenarios
it is assumed that a rapidly expanding and cooling thermalized system
experiences at the late stage of its evolution the so-called chemical
freeze-out, where all inelastic processes have to cease, accompanied
by the thermal freeze-out, which occurs when the mean free path of 
particles exceeds the linear sizes of the system. Therefore, the
conditions of the system at the chemical freeze-out stage can be 
obtained from hadron abundances and ratios, which are not affected by 
the collective flow. The analysis of experimental data taken in a 
broad energy range from SIS to SPS suggests \cite{CR99} that there 
exists a scaling law concerning the energy per hadron ratio at the
chemical freeze-out, $\langle E \rangle / \langle N \rangle \approx 1$ 
GeV, while in Au+Au collisions at $\sqrt{s} = 130$\,{\it A\/}GeV 
(RHIC) this ratio increases to $\langle E \rangle / \langle N \rangle 
\approx 1.1$ GeV. Simple estimates
show that in the latter case the average hadron mass should increase
as well, because the chemical freeze-out temperature at RHIC
does not exceed that at SPS by more than 10-20 MeV \cite{BMMRS01}.
In the present paper we argue that the rise of the average hadron 
energy and mass at RHIC is caused by the transition to dense 
meson-resonance rich matter with a considerable fraction of baryons
and antibaryons. Another substance with peculiar characteristics 
is baryon-dominated resonance matter, which can be formed in
heavy-ion collisions at bombarding energies between 10\,{\it A\/}GeV
and 40\,{\it A\/}GeV, accessible for the accelerator planned at GSI. 
Here the energy per hadron at chemical freeze-out increases as well. 

For our study we use a conventional statistical model (SM) of an
ideal hadron gas 
\cite{CR99,BMMRS01,BMHS99,LB56,Raf91,Bec96,Sol97,CEST97,YG99}
which enables one to determine  all macroscopic characteristics of a 
system at given temperature $T$, baryon chemical potential $\mu_B$,
and strangeness chemical potential $\mu_S$ via the set of 
distribution functions (in units of $c = k_B = \hbar =1$)
\beq 
\ds 
f(p,m_i) = \left[ \exp{\left( \frac{E_i - \mu_i}{T} \right) } 
\pm 1 \right] ^{-1} 
\label{eq1} 
\eeq
Here $m_i$, $E_i = \sqrt{p^2 + m_i^2}$, $p$, and $\mu_i$ are the mass,
energy, momentum, and the chemical potential of hadron species $i$,
respectively. Denoting baryon and strange charges of the $i$th 
particles as $B_i$ and $S_i$ one can write $\mu_i = B_i \mu_B + 
S_i\mu_S$. The electrochemical potential considered in \cite{CEST97} 
and the isospin chemical potential considered in \cite{BMMRS01,BMHS99}
are neglected. The sign ''$+(-)$" in Eq.~(\ref{eq1}) corresponds to 
fermions (bosons). The particle number density $n_i$, the energy 
density $\varepsilon_i$ and the partial pressure $P_i$ read
\beqar 
\ds 
\label{eq2}
n_i &=&\frac{g_i}{(2\pi)^3}\int_0^{\infty}f(p,m_i)
d^3p\ ,\\
\label{eq3} 
\varepsilon_i &=& \frac{g_i}{(2\pi)^3}\int_0^{\infty}
\sqrt{p^2+m_i^2}\, f(p,m_i) d^3p\ ,\\
\label{eq4}
P_i &=& \frac{g_i}{(2\pi)^3}\int_0^{\infty}
\frac{p^2}{3(p^2+m_i^2)^{1/2}} f(p,m_i) d^3p\ ,
\eeqar
with $g_i$ being the spin-isospin degeneracy factor of hadron $i$.
Instead of evaluating of the integrals in Eqs.~(\ref{eq2})-(\ref{eq4})
by replacing of Fermi-Dirac or Bose-Einstein distribution functions 
(\ref{eq1}) to Maxwell-Boltzmann ones
\beq
\ds
f^{MB}(p,m_i) = \exp{\left( \frac{\mu_i - E_i}{T} \right) }
\label{eq5} 
\eeq
we employ the series expansion of (\ref{eq1}) in a form \cite{LB56}
\beq
\ds
f(p,m_i) = \sum \limits ^{\infty}_{n=1} (\mp 1)^{n+1} \exp{\left(- n
\frac{E_i - \mu_i}{T} \right) } \ ,
\label{eq6}
\eeq
which is inserted to Eqs.~(\ref{eq2})-(\ref{eq4}).
After some straightforward calculations one gets
\beqar 
\ds 
n_i &=&\frac{g_i m_i^2 T}{2\pi^2}\sum_{n=1}^{\infty}
\frac{(\mp 1)^{n+1}}{n} \exp{ \left( \frac{n \mu_i}{T} \right)} 
K_2\left(\frac{n m_i}{T} \right) \\
\label{eq7}
\varepsilon_i &=& \frac{g_i m_i^2 T^2}{2\pi^2}\sum_{n=1}^{\infty}
\frac{(\mp 1)^{n+1}}{n^2} \exp{ \left( \frac{n \mu_i}{T} \right)} \\
\nonumber
&&\times \left[ 3 K_2\left(\frac{n m_i}{T} \right) + \frac{n m_i}{T} 
K_1\left(\frac{n m_i}{T} \right) \right] \\
\label{eq8} 
P_i &=& \frac{g_i m_i^2 T^2}{2\pi^2}\sum_{n=1}^{\infty}
\frac{(\mp 1)^{n+1}}{n^2} \exp{ \left( \frac{n \mu_i}{T} \right)} 
K_2\left(\frac{n m_i}{T} \right)
\label{eq9}
\eeqar
where $K_1$ and $K_2$ are modified Hankel function of first and
second order. For $n = 1$ the results with the Maxwell-Boltzmann 
distribution (\ref{eq5}) are regained \cite{Gro80}.

\begin{figure}[hbt]
\vspace{-1.5cm}
\includegraphics*[width=\linewidth]{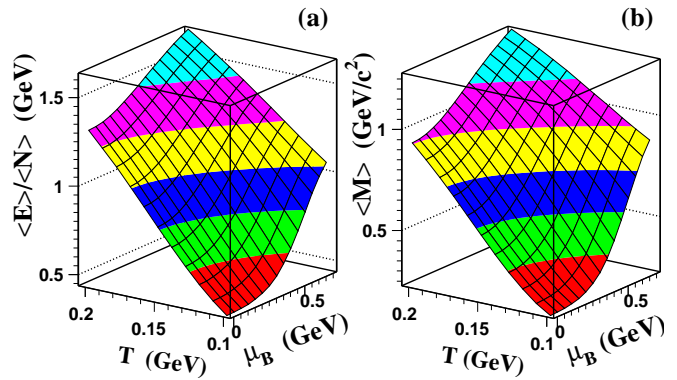}
\vspace{-1.0cm}
\caption{\label{fig1} (a) Average energy per hadron and (b) average 
hadron mass as functions of the temperature $T$ and the baryo-chemical 
potential $\mu_B$ of the system.}
\end{figure}

The equation of state (EOS) of an ideal hadron gas in the form
$\langle E(T, \mu_B) \rangle / \langle N(T, \mu_B) \rangle $ and the 
average hadron mass $\ds \langle M (T, \mu_B) \rangle = 1/N 
\sum_{i=1}^N m_i$ are shown in Fig.~\ref{fig1}. The calculations are 
performed within the standard SM 
at zero net strangeness in the system. The increase of both quantities
can be caused either by the rise of the temperature of the system or 
by an increasing baryon chemical potential, which is directly linked 
to the baryon density. Both macroscopic \cite{BMMRS01} and microscopic 
\cite{cell} models indicate that the case with low $\mu_B$ and 
relatively high $T$ is relevant for Au+Au collisions at RHIC. To study 
the conditions of chemical freeze-out we also apply a generalization 
of the SM to two thermal sources (TSM) proposed in \cite{tsm}, which 
accounts for possible inhomogeneities of the net baryon charge,
observed experimentally e.g. at SPS energies (158 AGeV/$c$) 
\cite{na52}, and net strangeness distribution inside the reaction 
volume. The idea behind the TSM is quite simple. It is well known
(see, e.g., \cite{CR99}) that the hadron ratios in the system 
consisting of several fireballs are not affected by longitudinal or
transverse collective motion of sources, provided all fireballs 
have the same baryon chemical potential and the same temperature.
In this case the particle ratios are identical to those obtained for 
a single static source. If, for instance, the baryon charge or
strangeness is not homogeneously distributed within the total
volume, the problem cannot be reduced to a single source scenario.
In the TSM the whole system is separated in two parts, which at SPS
and lower energies can most naturally be
interpreted as an inner source (or core) and an outer source (or halo), 
each being in local thermal and chemical equilibrium, i.e. their 
(local) macroscopic characteristics are determined by Eqs. 
(\ref{eq2}) - (\ref{eq4}). However, their temperatures, baryon and 
strangeness chemical potentials are allowed (but not postulated) to be 
different. No additional constraints are assumed in the
model except of the total strangeness conservation 
$N_S^{tot} = N_S^{(1)} + N_S^{(2)} = 0$. 
 
\begin{figure}[hbt]
\includegraphics*[width=\linewidth]{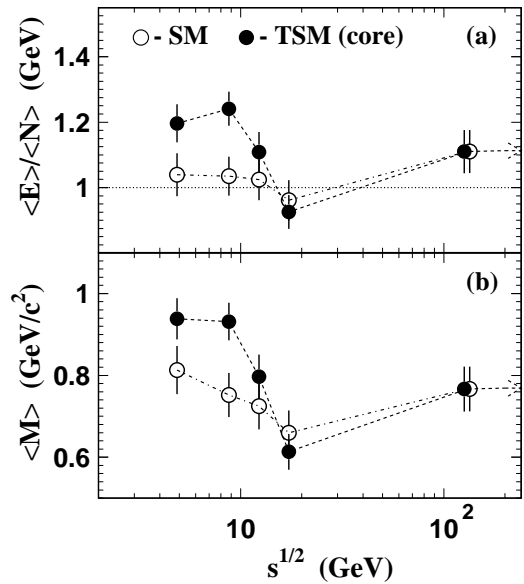}
\caption{\label{fig2} (a) Average energy per hadron and (b) average
hadron mass as functions of the center-of-mass energy of colliding
nuclei. Full circles correspond to central source of the TSM, open
circles denote the SM results. Lines are drawn to guide the eye.
}
\end{figure}

A fit to experimental data at SPS and RHIC energies has been performed 
in \cite{tsm}. Here it turns out that at RHIC the 
results of the fit to the SM and the TSM are identical: the TSM simply 
splits the volume of the system in two equal parts with similar 
characteristics, which means that the particles are really emitted
from one thermalized and homogeneous source. This gives us a 
temperature of $T = 176$\,MeV and a baryo-chemical potential of
$\mu_B = 39.8$\, MeV \cite{rem1}. The energy per hadron and the 
average hadron mass are $\langle E \rangle / \langle N \rangle = 1.12$ 
and $ \langle M \rangle = 0.77$\, GeV, respectively. As seen in 
Fig.~\ref{fig2}, this is a significant rise compared to the values of 
$\langle E \rangle / \langle N \rangle$ and  $ \langle M \rangle$ 
obtained from the fit to experimental data on heavy-ion collisions 
at lower energies. 

In contrast to RHIC, at lower energies results of the SM and the
TSM do not coincide anymore, although both models seem to provide 
reasonable agreement with the data, as shown in Table \ref{tab1}. 
\begin{table}
\caption{
\label{tab1}
Hadron yields and ratios for central heavy-ion collisions at
11.6{\it A} GeV, 40{\it A} GeV, and 80{\it A} GeV, respectively,
and results of the fit to the two-source and the single-source
statistical models of an ideal hadron gas.
}
\vspace*{1ex}
\begin{ruledtabular}
\begin{tabular}{lcccc}
                       & Data             &  TSM  &  SM   & Ref.\\
\hline
Au+Au  11.6{\it A} GeV &                  &       &       & \\         
$N_B$                  & 363$\pm 10 $     & 364.1 & 362.9 &
\protect\cite{e802_1} \\
$p/\pi^+$              & 1.234$\pm 0.126 $& 1.20  & 1.25  &
\protect\cite{e802_2} \\
$\pi^+$                & 133.7$\pm 9.9   $& 131.6 & 123.3 &
\protect\cite{e802_3} \\
$K^+$                  & 23.7$\pm 2.9    $& 25.77 & 28.55 &
\protect\cite{e802_1} \\
$K^-$                  & 3.76$\pm 0.47   $& 3.725 & 3.824 &
\protect\cite{e802_1} \\
$\Lambda$              & 20.34$\pm 2.74  $& 17.30 & 19.05 &
\protect\cite{ags_4} \\
$\bar{p}$              & 0.0185$\pm 0.002$& 0.0185& 0.0183&
\protect\cite{e802_5} \\
\hline
Pb+Pb  40{\it A} GeV   &                  &       &       & \\         
$N_B$                  & 349$\pm 5  $     & 351.6 & 352.8 &
\protect\cite{na49_1} \\
$\bar{\Lambda}/\Lambda$&0.025 $\pm 0.0023$& 0.0258& 0.0233&
\protect\cite{na49_3} \\
$\Lambda/p$ (Pb+Au)    & 0.22 $\pm 0.05  $& 0.212 & 0.232 &
\protect\cite{na45_1} \\
$\pi^-$                & 312  $\pm 15    $& 306.2 & 264.4 &
\protect\cite{na49_1} \\
$\pi^+$                & 282  $\pm 15    $& 276.2 & 239.0 &
\protect\cite{na49_1} \\
$K^+$                  &  56 $\pm 3      $& 57.4  & 63.2  &
\protect\cite{na49_1} \\
$K^-$                  & 17.8$\pm 0.9    $& 18.2  & 19.4  &
\protect\cite{na49_2} \\
$\bar{\Lambda}$        & 0.71 $\pm 0.07  $& 0.727 & 0.71  &
\protect\cite{na49_3} \\
\hline
Pb+Pb  80{\it A} GeV   &                  &       &       & \\         
$N_B$                  & 349$\pm 5  $     & 351.6 & 352.8 &
\protect\cite{na49_1} \\
$\pi^-$                & 445  $\pm 22    $& 403.7 & 366.7 &
\protect\cite{na49_4} \\
$\pi^+$                & 414  $\pm 22    $& 375.6 & 341.1 &
\protect\cite{na49_4} \\
$K^+$                  &  79 $\pm 5      $& 83.0  & 87.5  &
\protect\cite{na49_4} \\
$K^-$                  &  29 $\pm 2      $& 33.4  & 35.4  &
\protect\cite{na49_4} \\
$     \Lambda $        & 47.4 $\pm 3.7   $& 34.5  & 35.7  &
\protect\cite{na49_3} \\
$\bar{\Lambda}$        & 2.26 $\pm 0.1   $& 2.26  & 2.26  &
\protect\cite{na49_3} \\
\end{tabular}
\end{ruledtabular}
\end{table}
Here it is worth noting that the TSM analysis has been performed
for RHIC data taken in a very narrow midrapidity window. However,
results of the TSM fitted separately to merely $4\pi$-data and to
midrapidity data, obtained in Pb+Pb collisions at SPS energies,
indicate \cite{tsm} that the temperature and the baryon chemical
potential of the central source are varying insignificantly.
Therefore, we do not expect considerable changes of the core
conditions when the $4\pi$-data at RHIC will be available.
Macroscopic characteristics of the system obtained from the fit to
the models are listed in Table \ref{tab2}.
\begin{table*}
\caption{
\label{tab2}
Temperature $T$, baryon chemical potential $\mu_B$, strangeness
chemical potential $\mu_S$, net baryon density $\rho_B$, and net 
strangeness density $\rho_S$ obtained from the 
statistical model fit to experimental data on A+A collisions at 
11.6{\it A} GeV, 40{\it A} GeV, and 80{\it A} GeV. Of each three 
numbers the upper one corresponds to the single-source model, the 
middle number to the central core, and the lower one to the halo 
in the two-source model.
}
\vspace*{1ex}
\begin{ruledtabular}
\begin{tabular}{clllll}
    Reaction & $T$ (MeV) & $\mu_B$ (MeV) & $\mu_S$ (MeV) & 
    $\rho_B$ (fm$^{-3}$) & $\rho_S$ (fm$^{-3}$)  \\
\hline
                         & 123$\pm$5   & 558$\pm$15 & 122$\pm$4 & 
       0.149$\pm$0.01 & 0                           \\ 
 Au+Au, 11.6{\it A} GeV  & 141.4$\pm$5 & 564$\pm$15 & 142$\pm$4 &
       0.380$\pm$0.03 &-0.008$\pm 1\times 10^{-3}$  \\
                         & 100.6$\pm$7 & 558$\pm$17 &  94$\pm$3 & 
       0.004$\pm 3\times 10^{-4}$ & 0.0004$\pm 5\times 10^{-5}$  \\
\hline
                         & 148$\pm$5   & 367$\pm$14 & 84$\pm$3  &
       0.142$\pm$0.01 & 0                           \\ 
 Pb+Pb, 40{\it A} GeV    & 166$\pm$5   & 413$\pm$15 & 117$\pm$5 &
       0.430$\pm$0.03 &-0.0068$\pm 1\times 10^{-3}$ \\
                         & 103$\pm$7   & 352$\pm$15 &  35$\pm$3 & 
       0.007$\pm 5\times 10^{-4}$ & 0.0004$\pm 5\times 10^{-5}$  \\ 
\hline
                         & 155$\pm$5   & 284$\pm$15 &  66$\pm$2 &
       0.120$\pm$0.01 & 0                           \\ 
 Pb+Pb, 80{\it A} GeV    & 162.5$\pm$5 & 296$\pm$15 &  74$\pm$2 &
       0.190$\pm$0.015&-0.003$\pm 4\times 10^{-4}$  \\
                         &  98.3$\pm$7 & 313$\pm$15 &28.5$\pm$2 & 
       0.003$\pm 3\times 10^{-4}$ & 0.0005$\pm 5\times 10^{-5}$  \\
\end{tabular}
\end{ruledtabular}
\end{table*}
For A+A collisions at $E_{lab} = 11.6$\, 
AGeV, 40\,AGeV, and 80\,AGeV the energy per hadron and the average 
hadron mass in the central source (using the TSM) are about 20\% 
larger than those obtained in the SM because of the following reasons: 
At such energies the TSM favours the formation of hot and dense core 
surrounded by a cooler halo. Moreover, the model indicates that the 
net baryon charge and the net strangeness are non-uniformly 
distributed within the system volume. Here three points should be 
mentioned: Using the TSM, all antibaryons are contained in the core, 
which is in accord with experimental data on, e.g., $\bar{p}$, 
$\bar{d}$, and $\bar{\Lambda}$ production \cite{na52,na49_3,na49pap}. 
The net strangeness in the core is small and negative. Such a scenario 
is confirmed by microscopic ultra-relativistic quantum molecular 
dynamics (UrQMD) \cite{urqmd} model predictions \cite{lv99}. The 
negative strangeness in the central zone of a heavy-ion collision at 
energies below RHIC can be explained by different interaction cross 
sections of kaons and antikaons with baryons. In contrast, at RHIC 
energies and above the medium is meson dominated 
and the antibaryon yield at midrapidity is close to that of baryons. 
Hence, differences in the interaction cross sections are not 
important here leading to a homogeneous strangeness distribution over
the whole reaction volume. Finally, the SM indicates that the net 
baryon density of the fireball formed in A+A collisions at 11.6{\it A} 
GeV and 40{\it A} GeV is about 10\% below the normal nuclear density 
$\rho_0 = 0.17$\,fm$^{-3}$. At these energies hydrodynamic model 
calculations with and without the QGP formation \cite{Brprc94} and 
microscopic model simulations \cite{Brqm93,lv99} favor, however, the 
formation of a much denser matter with central baryon densities of
$\rho \approx 2$-$3\,\rho_0$ at chemical freeze-out, which should take 
place around $t = 7$-$10$\,fm/$c$. Thus, the TSM predictions of a 
small central source with baryon densities of $\rho_B \cong 2.5\,
\rho_0$ quantitatively agree with these estimates. The volume of the 
central source varies from 300\,fm$^3$ at AGS to 650\,fm$^3$ at 
40{\it A} GeV \cite{rem2}. Therefore, to probe this zone one has to 
study hadron abundances and ratios in a quite narrow midrapidity 
window. 

\begin{figure}[hbt]
\includegraphics*[width=\linewidth]{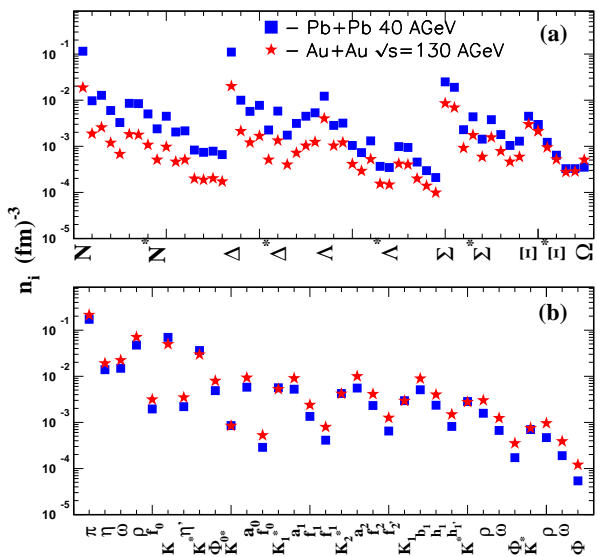}
\caption{\label{fig3} Densities of (a) baryons and (b) mesons in the
central zone of A+A collisions at $\sqrt{s}=130${\it A} GeV (stars)
and at 40{\it A} GeV (boxes) at chemical freeze-out (without feeding).
}
\end{figure}

The baryon fraction in the hadrons yield is decreasing from 64.5\% 
(59.3\%) in the TSM (SM) at AGS to 13.7\% at RHIC, while the 
antibaryon fraction rises from zero to 9.2\%. Figure~\ref{fig3}
depicts the hadronic densities at the chemical freeze-out 
stage in the central zone of heavy-ion collisions at
$\sqrt{s} = 130${\it A} GeV and at 40{\it A} GeV. The most abundant
species are $\pi$, $K$, $\rho$, $\bar{K}$, $\eta$, and
$\omega$ in the mesonic sector, and $N$, $\Delta$, $\Sigma$,
and $\Lambda$ in the baryonic one. Resonances are playing an important 
role in the hadronic spectra at both energies: the average baryon and
meson masses are 1.22(1.38) GeV and 432(585) MeV, respectively, at
40($\sqrt{s}=130$){\it A} GeV. 

The reason for the dropping of the average hadron mass and the energy 
per hadron in Pb+Pb at SPS and their subsequent rise in Au+Au 
collisions at RHIC is as follows: If the matter is dominated by 
baryons, then, despite the lower temperature at chemical freeze-out, 
the ratio $\langle E \rangle / \langle N \rangle$ is larger than 
1\,GeV. Compared to 
lower energies, the average baryon and meson masses at SPS are rising 
to $\langle M_B \rangle = 1.29$\, GeV and $\langle M_M \rangle = 470$ 
\, MeV, but the baryon+antibaryon fraction drops to 23\% of the total 
amount of hadrons. Therefore, the average mass of a hadron at chemical 
freeze-out decreases, as was first shown in \cite{CR99}. At RHIC 
energies the combined yield of baryons and antibaryons is essentially 
the same as that at SPS, but the average meson and baryon masses and 
kinetic energies increase due to the rise of the chemical freeze-out 
temperature. This leads to the rise of the energy per hadron.
Note, that if the freeze-out temperature increases insignificantly
with the rise of energy to $\sqrt{s} = 200${\it A} GeV, the combined 
$B + \bar{B}$ yield remains pretty stable when the baryon chemical 
potential $\mu_B$ drops to 29\,MeV, as predicted in \cite{BMMRS01},
and even to zero. Although in the later case the net baryon density
is essentially zero, the baryon+antibaryon yield still amounts
22.8\% of the total number of hadrons. Thus, the average energy per 
hadron and the average hadron mass should saturate at values $\langle 
E \rangle / \langle N \rangle \cong 1.1$\,GeV and $\langle M \rangle 
\cong 770$\,MeV.

In summary, average masses $\langle M \rangle$ and average energies
per hadron $\langle E \rangle / \langle N \rangle$ at chemical 
freeze-out have been analyzed for heavy-ion collisions at energies 
spanning from AGS to RHIC. It is found that the rise of the 
$\langle E \rangle / \langle N \rangle$ ratio
from 1\,GeV to 1.1\,GeV at RHIC is caused by the formation of a hot 
resonance-rich substance, in which the fraction of antibaryons 
becomes significant. Despite of the low net baryon density, the 
combined relative yield of baryons and antibaryons is comparable to
that at SPS, while the average masses of both baryons and mesons 
are increasing due to a higher freeze-out temperature.
If the temperature of chemical freeze-out will not exceed the limit 
of 180\,MeV with rising incident energy, the saturation
values are $\langle E \rangle / \langle N \rangle = 1.1$\,GeV and 
$\langle M \rangle = 770$\,MeV even at zero baryo-chemical potential.
A possible violation of the scaling $\langle E \rangle / \langle N 
\rangle \cong 1$\,GeV in A+A collisions at energies between 
11.6{\it A} GeV and 40{\it A} GeV has been discussed. We show 
that in baryon-dominated matter with baryon densities of 2-3 normal 
baryon density the energy per hadron ratio should rise to about 
1.2\,GeV.  This estimate can be checked by measuring hadron yields 
and ratios in a very narrow range at midrapidity in heavy-ion 
reactions at energies of the planned GSI accelerator.

We are grateful to L.P. Csernai, M. Gazdzicki, M. Gorenstein, 
K. Redlich, D. Strottman, and N. Xu for the fruitful discussions.
The work was supported by the
Deutsche Forschungsgemeinschaft (DFG), Bundesministerium f\"ur
Bildung and Forschung (BMBF) under the contract No. 06T\"U986, and
the National Science Foundation of China under the contracts
No. 19975075 and No. 19775068.




\begin{thebibliography}{9}

\bibitem{QM99} Proceedings of the QM'99 conference
(Torino, Italy, 1999)
[Nucl. Phys. {\bf A661}, 1c (1999)].

\bibitem{QM01} Proceedings of the QM'2001 conference
(Stony Brook, NY, USA, 2001)
[Nucl. Phys. {\bf A698}, 1c (2002)].

\bibitem{CR99}  J.~Cleymans and K.~Redlich,
Phys. Rev. C {\bf 60}, 054908 (1999); Phys. Rev. Lett {\bf 81},
5284 (1998).

\bibitem{BMMRS01} P.~Braun-Munzinger, D.~Magestro, K.~Redlich, and
J.~Stachel, Phys. Lett. B {\bf 518}, 41 (2001).

\bibitem{BMHS99} P.~Braun-Munzinger, I.~Heppe, and J.~Stachel,
Phys. Lett. B {\bf 465}, 15 (1999).

\bibitem{LB56} L.D.~Landau and S.Z.~Belenkij,
Usp. Phys. Nauk {\bf 56}, 309 (1955) [Nuovo Cim. Suppl. {\bf 3},
15 (1956)].

\bibitem{Raf91} J.~Rafelski, Phys. Lett. B {\bf 62}, 333 (1991);
J.~Letessier and J.~Rafelski, Phys. Rev. C {\bf 59}, 947 (1999).

\bibitem{Bec96} F.~Becattini, Z. Phys. C {\bf 69}, 485 (1996);
F.~Becattini and U.~Heinz, Z. Phys. C {\bf 76}, 269 (1997);
F.~Becattini, M.~Gazdzicki, and J.~Sollfrank,
Nucl. Phys. {\bf A638}, 403 (1998).

\bibitem{Sol97} J.~Sollfrank, J. Phys. G {\bf 23}, 1903 (1997).

\bibitem{CEST97} J.~Cleymans, D.~Elliot, H.~Satz, and R.L.~Thews,
Z. Phys. C {\bf 74}, 319 (1997).

\bibitem{YG99} G.D.~Yen and M.I.~Gorenstein,
Phys. Rev. C {\bf 59}, 2788 (1999).

\bibitem{Gro80} S.R. de Groot, W.A. van Leeuwen, and Ch.G. van Weert,
{\it Relativistic Kinetic Theory\/} (North Holland, Amsterdam, 1980).

\bibitem{cell} L.V.~Bravina {\it et al.\/},
Nucl. Phys. {\bf A698}, 383c (2002); 
Phys. Rev. C {\bf 63}, 064902 (2001);
J. Phys. G {\bf 27}, 421 (2001).

\bibitem{tsm} Z.-D.~Lu, A.~Faessler, C.~Fuchs, and E.E.~Zabrodin,
nucl-th/0110040;
Talk at Strangeness'2001 conference (Frankfurt a.M., Germany, 2001), 
[J. Phys. G. (in press)].

\bibitem{na52} R.~Arsenescu {\it et al.\/}, NA52 Collab.,
J. Phys. G {\bf 25}, 225 (1999).


\bibitem{rem1} Compared to \cite{tsm}, the multiplicity of 
negatively chagred hadrons has been excluded from the present fit. 
With negatively charged hadrons the temperature and the baryon
chemical potential rises to 185 MeV and 52.5 MeV, correspondingly.

\bibitem{e802_1} L.~Ahle {\it et al.}, E802 Collab.,
Phys. Rev. C {\bf 60}, 044904 (1999). 

\bibitem{e802_2} L.~Ahle {\it et al.}, E802 Collab.,
Phys. Rev. C {\bf 60}, 064901 (1999).

\bibitem{e802_3} L.~Ahle {\it et al.}, E802 Collab.,
Phys. Rev. C {\bf 59}, 2173 (1999). 

\bibitem{ags_4} S.~Ahmad {\it et al.}, 
Phys. Lett. B {\bf 382}, 35 (1996).

\bibitem{e802_5} L.~Ahle {\it et al.}, E802 Collab.,
Phys. Rev. Lett. {\bf 81}, 2650 (1998). 

\bibitem{na49_1} C. Blume {\it et al.\/}, NA49 Collab., 
Nucl. Phys. {\bf A698}, 104c (2002). 

\bibitem{na49_3} A. Mischke {\it et al.\/}, NA49 Collab., 
nucl-ex/0201012.

\bibitem{na45_1} K. Filimonov {\it et al.\/}, CERES/NA45 Collab.,
nucl-ex/0109017.

\bibitem{na49_2} V. Friese {\it et al.\/}, NA49 Collab., 
Nucl. Phys. {\bf A698}, 487c (2002). 

\bibitem{na49_4} T. Kollegger {\it et al.\/}, NA49 Collab.,
nucl-ex/0201019.

\bibitem{na49pap} J.~B\"achler {\it et al.\/}, NA49 Collab.,
Nucl. Phys. {\bf A661}, 45c (1999).

\bibitem{urqmd} S.A.~Bass {\it et al.\/}, 
Prog. Part. Nucl. Phys. {\bf 41}, 255 (1998);
M.~Bleicher {\it et al.\/}, J. Phys. G {\bf 25}, 1859 (1999).

\bibitem{lv99} L.V.~Bravina {\it et al.\/},
Phys. Lett. B {\bf 434}, 379 (1998); J. Phys. G {\bf 25}, 351 (1999);
Phys. Rev. C {\bf 60}, 024904 (1999).

\bibitem{Brprc94} L.~Bravina, L.P.~Csernai, P.~Levai, and D.~Strottman,
Phys. Rev. C {\bf 50}, 2161 (1994).

\bibitem{Brqm93} L.V.~Bravina, N.S.~Amelin, L.P.~Csernai, P.~Levai, 
and D.~Strottman, Nucl. Phys. {\bf A566}, 461c (1994).

\bibitem{rem2} Note, that these numbers are obtained for an ideal gas
of pointlike hadrons. If the Van der Waals type of the EOS with the
non-zero hard core radius for hadrons is employed, the volume of the 
source increases. 

\end{thebibliography}
\end{document}